# End-Point Detection with State Transition Model based on Chunk-Wise Classification

*Juntae Kim[†], Jaesung Bae[†] and Minsoo Hahn*

School of Electrical Engineering,
Korea Advanced Institute of Science and Technology (KAIST), Daejeon, Korea

`{jtkim,bjsd3,mshahn2}@kaist.ac.kr`

## Abstract

A state transition model (STM) based on chunk-wise classification was proposed for end-point detection (EPD). In general, EPD is developed using frame-wise voice activity detection (VAD) with additional STM, in which the state transition is conducted based on VAD's frame-level decision (speech or non-speech). However, VAD errors frequently occur in noisy environments, even though we use state-of-the-art deep neural network based VAD, which causes the undesired state transition of STM. In this work, to build robust STM, a state transition is conducted based on chunk-wise classification as EPD does not need to be conducted in frame-level. The chunk consists of multiple frames and the classification of chunk between speech and non-speech is done by aggregating the decisions of VAD for multiple frames, so that some undesired VAD errors in a chunk can be smoothed by other correct VAD decisions. Finally, the model was evaluated in both qualitative and quantitative measures including phone error rate.

**Index Terms**: end-point detection, voice activity detection.

## 1. Introduction

End-point detection (EPD) is an important front-end for speech recognition systems. EPD detects the end-point of incoming utterances, enabling speech recognition systems to effectively reduce the search space at the decoding stage so that it can reduce the computation cost while expecting some performance improvement [1-5].

A classic EPD is based on frame-wise voice activity detection (VAD) with additional state transition model (STM) [1, 6-7]. In conventional STMs, the state changes from non-speech to speech when incoming frame is detected as speech by VAD. In that condition, if current state is speech state, when consecutive non-speech frames longer than a given threshold are detected, the state changes from speech to non-speech and STM decides the time at which state transition occurred, as end-point.

As state transition in STM occurs based on VAD's decision (speech or non-speech), the performance of VAD is crucial for EPD. Recently, deep-learning-based VADs using deep neural network (DNN) [8-9], recurrent neural network (RNN) [10-12] and convolutional neural network (CNN) [13] have been outperformed by the conventional VADs [14-16]. However, VAD error always can occur even in clean environment although we use state-of-the-art VAD, which causes undesired state transition of STM. Further, the EPD should not be conducted in frame-level. As each frame has much shorter duration than that of a phone for speech recognition, it is not necessary to frequently conduct state transition of STM whenever an incoming VAD's frame-level decision is updated.

In this study, a simple but powerful STM is proposed based on chunk-wise classification for EPD, referred to as CEPD. The state transition of proposed STM occurs according to the chunk-wise classification rather than the VAD's frame-wise decision. The chunk consists of multiple frames and the chunk-wise classification is conducted by aggregating the decisions of VAD for multiple frames within a chunk, so that some undesired VAD errors in a chunk can be smoothed by other correct VAD decisions. To verify CEPD, its performance in both qualitative and quantitative measures was investigated, including proposed metrics that can be related with speech recognition performance and phone error rate (PER). Although detecting the start-point as well as end-point was considered, conventional term, referred to as EPD was followed.

## 2. Proposed End-Point Detection System

The proposed EPD framework consists of two stages: VAD and STM. Firstly, frame-wise VAD based on DNN (DNN-VAD) is conducted. Then, decisions of VAD at frame-level are aggregated in a chunk-level for chunk-wise classification. The proposed STM finally obtains the start- and end-point of incoming utterance, depending on chunk-wise classification results. The next section described the details.

### 2.1. Voice activity detection based on deep neural network

The input feature vectors for DNN-VAD are extracted from input frames and represented as $\{v_m\}_{m=1}^{M}$, where $m = 1, 2, …, M$ is the frame number. Then, the decision of DNN-VAD $y_n \in \{0, 1\}$ is obtained as follows:

$$y_n = DNN(v_{n-\tau}, ..., v_n, ..., v_{n+\tau}) \quad (1)$$

where $n$ is the frame number and $\tau$ is an integer number. The details of DNN setup is described in section 3.1.2.

### 2.2. State transition model using chunk-wise classification

The chunk consists of $2w$ frames. The chunk-wise classification is conducted as follows:

$$C_i = \sum_{k=(i-1)\cdot w}^{(i+1)\cdot w - 1} y_k / 2w \quad (2)$$

---

[†] These authors contributed equally to this work.

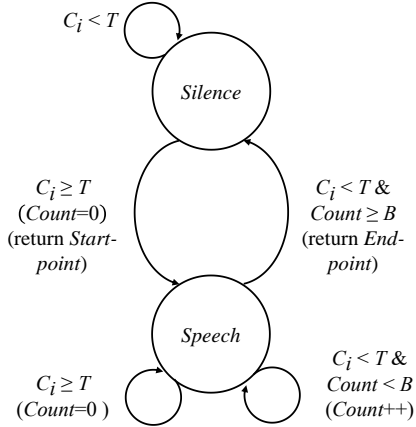

Figure 1: *STM of CEPD*

where $i$ is the chunk number, $y_k$ is the decision of $k$ th frame using DNN-VAD described in section 2.1 and $C_i$ is the soft decision of chunk-wise classification. Based on this chunk-wise classification, the STM is defined as described in Figure 1. The STM has two states: *silence* and *speech*. The state transition conditions are written with the transition arrows and the actions are described in parentheses. The $T \in [0,1]$, $B$, and *Count* are threshold, buffer, and chunk counter, respectively. The chunk counter counts the number of consecutive non-speech classified chunks and the buffer is the required number of consecutive non-speech chunks for end-point detection. In this study it is assumed that the *silence* state is the initial state.

The state transition from *silence* to *speech* occurs if $C_i \geq T$. While transition, the STM outputs the start-point corresponding to the first frame in the chunk, and sets the *Count* to zero.

During the *speech* state, the state is kept to *speech* when $C_i \geq T$ or when $C_i < T$ and *Count* $< B$ for which, the *Count* is re-initialized to zero or increase by one, respectively. By adopting the latter case, the immediate state transition from *speech* to *silence* can be prevented when $C_i < T$. Therefore, the STM can alleviate the division of speech utterances errors.

Finally, in the case of $C_i < T$ and *Count* $\geq B$, the *speech* state is changed to *silence* state and STM outputs the end-point corresponding to the last frame in the chunk.

In addition, the *minimum speech duration* and *maximum speech duration* conditions when state transition from *speech* to *silence* occurs was proposed. If the detected speech duration (from start-point to end-point) does not meet the conditions, the STM rejects the results and is initialized to *silence* state without outputting any start- and end-point. This condition will be useful in tasks where the expected length of speech is known and fixed. Section 3.1.4 contains detailed parameter settings of STM of CEPD.

## 3. Experiments

### 3.1. Experimental setup

#### 3.1.1. Evaluation dataset

For evaluation, TIMIT [17] test dataset containing 1344 utterances was adopted. The sampling rate of TIMIT dataset is 16 kHz. When framing, 10 and 25 ms window shift and size, respectively were used. As TIMIT utterances do not have enough silence before and after speech segments, which is not realistic in real applications, certain lengths of silence before and after speech segments were added, until the ratio of speech segments to length of utterance (SR) become 30 and 50. For e.g., if SR is set to 50, and the length of utterance 1 s, the speech segments will be 0.5 s. TIMIT ground truth labels of start- and end-point were used.

To verify the proposed method in the noisy environment, a noisy dataset was built by adding some noises to TIMIT test dataset. NOISEX-92 corpus [18] containing 15 types of noise: white, machinegun, babble, volvo, etc., was used for noise dataset. One of the noise types was randomly picked and was added to each silence-added utterance with signal-to-noise ratio (SNR) randomly selected between 10 and 20 dB. When adding the noise, FaNT tools were used [19]. In summary, 1344 × 4 = 5376 utterances were tested, including clean and noisy cases with SR at 30 and 50, respectively.

#### 3.1.2. Deep neural network based VAD setup

Log-power-spectra features (LPS) for DNN's input features were used. The FFT point for LPS was set to 512 so that feature dimension was 257. The input LPS were locally z-score normalized, i.e., z-score normalization was conducted per incoming chunk. To utilize the context information, 5 past and future frames were spliced, $\tau = 5$, to the current frame so that 11 frames were used as input frames, corresponding to 11 × 257 = 2827 input features for DNN. The DNN had two hidden layer and the number of hidden units were 512. It has rectified linear unit (ReLU) [20] activation function. The training loss function for DNN was set to typical cross entropy loss. For training, Adam [21] optimizer was used and the optimal initial learning rate was obtained by a random search method [22]. The batch size was set to 4096, and batch normalization was applied [23]. Dropout with rate of 0.5 was also used [24].

To train DNN-VAD, TIMIT train dataset was adopted; 95% of utterances were used as training and remaining 5% were used as validation. As TIMIT utterances have silence duration much shorter than speech, 2-s-long silence segments were added before and after each utterance [25]. The ground truth labels in TIMIT were used for VAD's true labels. For training noise dataset, sound effect library [26] was adopted for unseen noise scenarios. Approximately 5000 sound effects were randomly extracted from that library for building the noisy utterances. These extracted sound effects were concatenated into long sound wave. After that, an utterance was randomly picked from the silence-added TIMIT train dataset and added to the long sound wave with a certain SNR in -10 to 12 dB; this procedure was repeated until the end of long sound wave. Relatively high SNRs were used at the initial training and then gradually decreased SNR level as underfitting problems were identified if harsh noise was introduced at the early stage of training.

DNN-VAD described in this section was used for both baseline and proposed STM to build EPD.

#### 3.1.3. Baseline state transition model

The baseline STM uses the frame-wise VAD decision $y_k$ for state transition. As the shifting size of chunk is $w$ times larger than the frame, the buffer size $wB$ was used instead of $B$, so that the buffer sizes for baseline and proposed STM in time domain were set equal for fair comparison. We refer baseline method as FEPD.

Table 1: *Performance comparison on silence-added TIMIT test dataset with SR 30 and 50 over clean and noisy environment. For ES, LS, EE and LE, median values were obtained from corresponding utterances. The '-' in LS means that there is no utterance corresponding to LS. The PER was averaged over 1344 utterances for each case. The numbers in bold indicate the best results.*

| SR | Method | Noise | NDU | DU | ES (ms) | LS (ms) | EE (ms) | LE (ms) | PER (%) |
|---|---|---|---|---|---|---|---|---|---|
| 30 | FEPD | Clean | 1109 | 13 | 227 | - | **28** | 520 | 28.50 |
|  |  | Noisy | 4573 | **14** | 153 | 29 | 52 | 430 | 37.27 |
|  | CEPD | Clean | **3** | **11** | 68 | 15 | 46 | **401** | **26.89** |
|  |  | Noisy | **569** | 23 | 66 | 19 | 45 | 364 | **35.70** |
| 50 | FEPD | Clean | 354 | 14 | 229 | - | 31 | 517 | 28.27 |
|  |  | Noisy | 1645 | 23 | 156 | 20 | 47 | 428 | 35.87 |
|  | CEPD | Clean | **5** | **11** | 66 | 15 | 36 | **406** | **26.56** |
|  |  | Noisy | **205** | **20** | 66 | 18 | 44 | 363 | **35.10** |

Table 2: *PER comparison when EPD with ground truth start- and end-point (Oracle EPD) were applied and EPD was not used. (No EPD)*

| Type | SR | Noise | PER (%) |
|---|---|---|---|
| Oracle EPD | - | Clean | 26.14 |
|  |  | Noisy | 36.16 |
| No EPD | 30 | Clean | 41.04 |
|  |  | Noisy | 44.85 |
|  | 50 | Clean | 35.94 |
|  |  | Noisy | 39.25 |

### 3.1.4. Proposed state transition model setup

The proposed STM has the parameters: *B*, *w*, *min speech duration*, *max speech duration* and *T*. They were set to 5, 10, 500 ms, 10 s and 0.5 respectively. *B* and *W* were found from the validation set. Though the max and min speech duration were set to handle the exception, no exception occurred in the experiments. In addition, the *T* was not optimized according to the noise types for fair comparison with baseline STM.

### 3.1.5. Phone recognizer

To verify the proposed EPD's effect for the noise robust speech recognition task, the phoneme recognition experiments were conducted with the test dataset. For training dataset, we used the noisy dataset introduced in Section 3.1.2, with corresponding speech phone labels in TIMIT. Phone recognizer was used based on DNN with Hidden Markov model (DNN-HMM) in [27]. The DNN consists of 7 hidden layers with 2048 nodes and 8839 output units, corresponding to the number of tied tri-phone states. For the input features, 13 dimensional mel-frequency cepstral coefficients (MFCCs) were extracted from 11 frames, composed of one current frame and 5 left right frames of it. They were reduced into 40 dimensions by applying linear discriminant analysis (LDA) and z-score normalization. For training, the restricted Boltzmann machine based [28] was first conducted as pre-training and cross entropy loss based supervised training with alignment information from Gaussian mixture model with HMM (GMM-HMM) trained on clean speech signal was followed. The DNN was then re-trained based on sequence discriminative training [27].

### 3.1.6. Evaluation metrics

To validate EPD methods, following metrics were used: (i) Number of utterances detected in non-speech section (NDU): The utterances counted as NDU have only noise instead of speech segments. The NDU in ideal case is 0; (ii) Number of divided utterances (DU): As the division of original utterance could affect the performance in speech recognition task, DU was used as EPD's evaluation metric. If one utterance was detected into two utterances, that case was counted as DU. The DU in ideal case is 0; (iii) Early start-point error (ES): The ES is the difference between early detected start-point and ground truth start-point in duration (ms); (iv) Late start-point error (LS): The LS is the difference between the lately detected start-point and ground truth start-point in duration (ms); (v) Early end-point error (EE): The EE is the difference between early detected end-point and ground truth end-point in duration (ms); (vi) Late end-point error (LE): The LE is the difference between lately detected end-point and ground truth end-point in duration (ms). Whenever ES, LS, EE, and LE were calculated, the utterances counted for DU and NDU were ignored.

### 3.2. Experimental results and discussion

Table 1 compares FEPD with CEPD, which outperformed FEPD in all metrics. The most outstanding difference between FEPD and CEPD was found in NDU. As described in Figure 2-(c), FEPD starts to detect an utterance whenever VAD classifies an input frame as speech because the state transition of STM in FEPD is conducted as per the VAD's frame-wise classification. Hence, it is quite sensitive to the VAD's noise, that is detected as speech (NDS) error. However, the state transition of CEPD is conducted as per chunk-wise classification, so that the state transition could be robust to sparsely occurred VAD's NDS error as described in Figure 2-(c) in which the soft decision of chunk-wise classification has smaller value in 0-1 s, though VAD causes NDS errors.

DU has trade-off relationship with NDU, as DU and NDU are sensitive to speech-detected as noise and NDS error, respectively. In spite of this relationship, CEPD still outperformed FEPD except for the noisy case in SR 30. Figure 2-(b), describes the case for which DU occurred when using CEPD. In that case, CEPD divided one utterance into two utterances as VAD sparsely decided frames in 4.2-4.7 s as speech, so that state transition of STM in CEPD was conducted to the silence state. DU should be carefully treated as it can degrade the performance of following speech phone recognition system owing to its inability to use some context information between divided utterances at the decoding stage. The possible simple solutions are: (i) Increasing the buffer size (*B*); and (ii) Adjusting the threshold (*T*) for state transition. Though the first solution deterministically increases EE, the second solution can be heuristic as the optimal threshold can be different according to noise types.

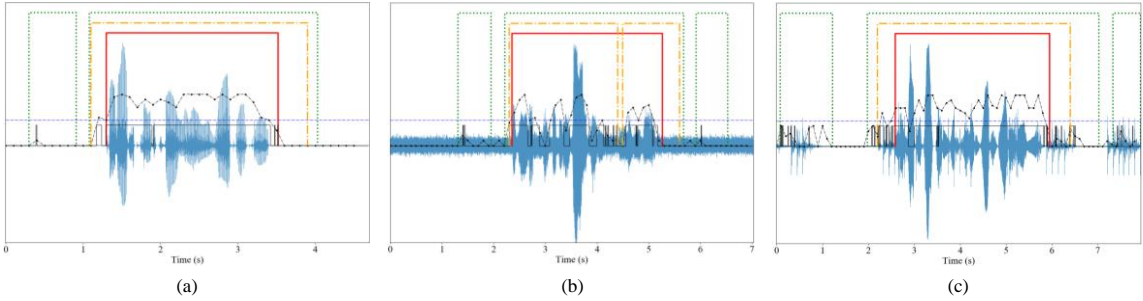

Figure 2: *Examples of EPD results at (a) clean, (b) white noise and (c) machinegun noise environment. All examples are in SR 50 condition and SNR is set to 10 dB for noisy environment. Each figure includes ground truth labels (solid red line), CEPD (orange dashed line), FEPD (dotted green line), frame-wise VAD decisions (solid black line), soft decision of chunk-wise classification (black dashed-dot line with black circle), and the threshold 'T' (blue dash-dot line).*

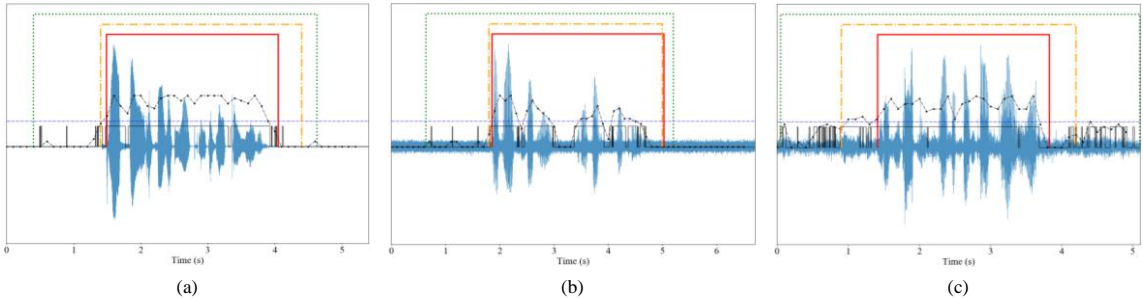

Figure 3: *Examples of EPD results at (a) clean, (b) white noise, and (c) babble noise environment. Other descriptions are same with Figure 2.*

In both ES and LE, CEPD still outperformed FEPD, although CEPD shows high LE in absolute. However, as in Figure 2- and 3-(a), VAD could not show perfect performance even in clean cases. Further, EPD should not detect the utterance in perfect, i.e. some silence segments before and after detected utterance can be allowed as phone recognition system individually has some method dealing with silence, e.g., silence symbol in DNN-HMM, and blank symbol in connectionist temporal classification (CTC) [29] based phone recognition systems. In Figure 3-(b) FEPD shows higher ES than CEPD even if the VAD was conducted relatively well. In Figure 3-(c), FEPD shows severe ES and LE, because of frequently occurred NDS errors while CEPD shows reasonable performance in SE and EE.

In LS and EE, CEPD could not outperform FEPD in great margin. LS and EE can be more crucial compared to ES and LE as front and back of speech signal can be lost by severe LS and EE. However, both LS and EE values of FEPD and CEPD are in acceptable range (< 50 ms), considering ground truth start- and end-point have some margin compared to real speech start- and end-point (> 100 ms). Further, we investigated that LS and EE occurred by FEPD and CEPD did not degrade PER, implying that both start- and end-point detected by FEPD and CEPD did not severely delete some front and back of speech signal. Apart from Table 1, it was also found that ES and LE occurred more frequently than LS and EE. The ratio between counted number of occurrences for LS and ES was approximately 1:747 and 1:6, for FEPD and CEPD, respectively. For EE and LE, the ratio was 1:119 and 1:43 for FEPD and CEPD, respectively.

To verify the effect of EPD to speech phone recognition task, PER was investigated. Table 2 describes the PER when ground truth start- and end-point were used for EPD (oracle EPD) and EPD not applied (no EPD). Table 2 clearly shows that applying EPD outperformed that of not applying EPD. This result is elementary as applying EPD reduces the search space of decoder in phone recognition system, while it also internally can handle some silence or noises. In Table 1, it can be seen that CEPD outperformed FEPD for PER in all cases, which is elementary as CEPD outperformed FEPD in SE and EE, related with size of search space of decoder in phone recognition system.

## 4. Conclusions

In this work, EPD was proposed including DNN-VAD and STM with chunk-wise classification. As the state transition of STM was conducted according to the result of chunk-wise classification, more noise robust EPD could be obtained compared to conventional one. It was also found that CEPD outperformed FEPD in almost all metrics including PER. For chunk-wise classification, the decisions of multiple frames using DNN-VAD were averaged. Although averaging method shows outstanding performance, this work shows that this method can be improved by adopting DNN based chunk-wise classification, which will be the future work of the authors.

## 5. Acknowledgements

This work was supported by the Basic Science Research Program through the National Research Foundation of Korea (Grant No. 2017R1A2B4011357) funded by the Ministry of Science, ICT, and Future Planning.